\documentstyle[11pt,newpasp,twoside,epsf]{article} 
\begin{document} 
\title{White Dwarfs in Globular Clusters
-- Progenitors, Successors and the Real Thing}
\author{Sabine Moehler} 
\affil{Institut f\"ur theoretische Physik und Astrophysik,
Abt. Astrophysik, D-24098 Kiel}

\begin{abstract} 
I will start by discussing the evolutionary status of the white dwarf
{\em progenitors}, the hot UV bright stars.
Observations of UIT-selected UV bright stars in globular
clusters suggest that a high percentage of them manage to evolve from
the horizontal branch to the white dwarf region without passing
through the thermally pulsing AGB phase, thereby avoiding the
planetary nebula stage.

The white dwarf {\em successors} are stars experiencing a very late
helium core flash while already on the helium white dwarf cooling
curve. While they have been around theoretically for quite some time
strong candidates could be verified only quite recently.

And, last not least, the {\em white dwarfs} themselves offer new
opportunities to derive distances and ages of globular clusters, which
I will discuss. For a discussion of white dwarfs in binaries see the
reviews by Adrienne Cool and Frank Verbunt in this volume.
\end{abstract}

Being a spectroscopist at heart I concentrate in this review mainly on
results that were obtained from spectroscopic analyses. However, as
you will see in the last section of this review there are stars that
manage to evade spectroscopic analyses even with the latest 8-10m
class telescopes.

\section{Progenitors: UV Bright Stars} 

UV bright stars have been classically defined as stars brighter than
the horizontal branch and bluer than the red giant branch (Zinn et
al.\ 1972). Such stars are also brighter in $U$ than any other cluster
star. UV bright stars are produced by evolution from the
\begin{itemize}
\vspace*{-1ex}
\item {\em horizontal branch} (HB) towards the asymptotic
giant branch (post-HB stars)
\vspace*{-1ex}
\item {\em extreme HB} (EHB, T$_{\rm eff} >$23,000~K) 
directly towards the white dwarf domain (post-EHB stars)
\vspace*{-1ex}
\item {\em asymptotic giant branch} towards the white dwarf domain (post-AGB
stars), possibly showing up as central stars of planetary nebulae.
\vspace*{-1ex}
\end{itemize}

Ground based searches like the one of Zinn et al. (1972) found
primarily cool UV bright stars (T$_{\rm eff}
\le$9000~K). Spectroscopic analyses of the few hot UV bright stars
found this way turned up only post-AGB stars. 
This finding is puzzling as a minimum mass of
0.565~M$_\odot$ is required to reach the thermally pulsing AGB
(Sch\"onberner 1983), which makes this evolutionary stage difficult to
reach for the low mass stars in today's globular clusters.  The
post-AGB phase also has a shorter lifetime than the other channels
producing UV bright stars, making the observation of such stars even
further unlikely.

\begin{figure}
\plotfiddle{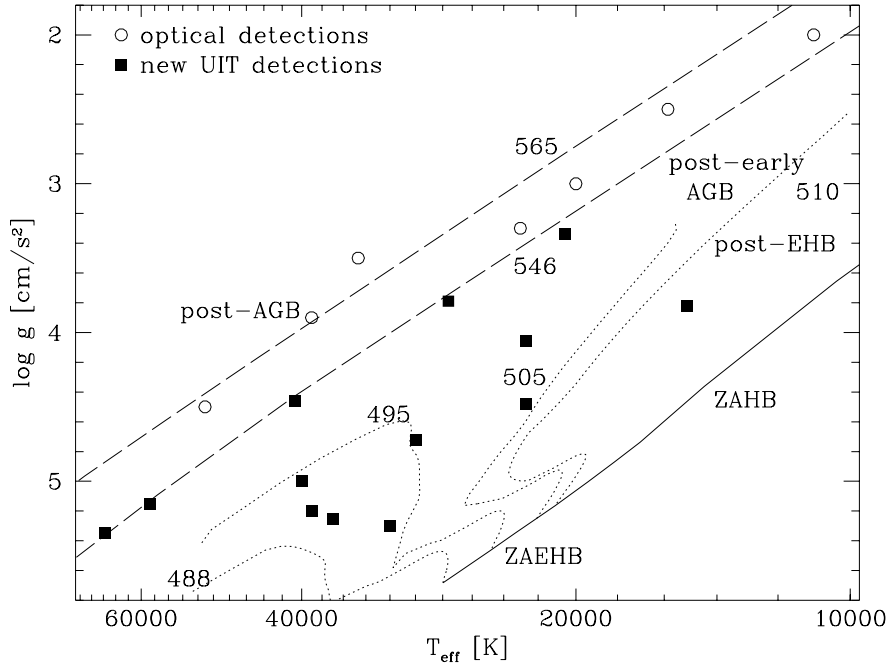}{8.3cm}{270}{45}{45}{-180}{255}
\caption{Results of spectroscopic analyses of UV bright stars compared
to evolutionary tracks. Open circles mark stars detected by optical
searches (Conlon et al. 1994;
Dixon et al. 1994, 1995; Glaspey et al. 1985, Heber \& Kudritzki 1986,
Moehler et al. 1998b, Rauch et al. 2002), filled squares mark UV
bright stars detected as such by UIT (Moehler et al. 1998a, Landsman
priv. comm.). The post-AGB and post-early AGB tracks are from
Sch\"onberner (1983), the zero-age HB/EHB (ZAHB/ZAEHB)
and post-EHB tracks are from Dorman
et al. (1993). The numbers give the stellar mass of the track in units
of 0.001~M$_\odot$.}
\end{figure}

However, optical searches are obviously biased against finding hot
stars, whose flux maximum moves ever farther to the ultraviolet with
increasing temperature. Therefore searches in the satellite
ultraviolet are much more promising and indeed the Ultraviolet Imaging
Telescope (Stecher et al. 1997) found quite a few new hot UV bright
stars in the globular cluster it surveyed. Spectroscopic analyses of
these stars by ground based (ESO, Moehler et al. 1998a) and HST (Landsman,
priv. comm.) observations showed stars evolving away from the extreme
HB and also post-early AGB stars, that left the AGB {\em before} the
thermal pulses started, but no new ``classical'' post-AGB stars (see
Moehler 2001 for more details). These new results are in much better
agreement with the expectations from stellar evolution theory with
respect to evolutionary life times (and thus observability) and minimum
masses. At the same time they also suggest an explanation for the {\em lack
of planetary nebulae} in globular clusters observed by Jacoby et
al. (1997): None of the UV bright stars detected solely by UIT will
produce a planetary nebula -- post-EHB stars never reach the AGB and
the evolution of post-early AGB stars proceeds so slowly that by the
time they are hot enough to excite the remnants of their AGB envelopes
these remnants have evaporated.

\section{Successors: Blue Hook Stars}

The globular clusters $\omega$~Cen and NGC~2808 show a large number of
hot horizontal branch (HB) stars that populate a very long blue tail down
to rather faint visual magnitudes.  Observations of $\omega$~Cen
(Whitney et al. 1998; D'Cruz et al. 2000) and NGC~2808 (Brown et
al. 2001) in the far-UV revealed a puzzling feature: At the very hot
end the HB shows a spread in UV brightness that cannot be explained by
measuring errors.  While stars brighter than the zero-age HB (ZAHB)
might be understood by evolution away from the ZAHB, the stars fainter
than the ZAHB cannot be explained by canonical HB evolution. The
fainter stars appear to form a hook-like feature extending up to 0.7
mag below the ZAHB and are therefore called ``blue hook'' stars.

Classical HB stars burn helium in a core of about 0.5M$_\odot$ and
hydrogen in a shell. They have hydrogen-rich envelopes of varying mass
and the more massive the envelope, the cooler is the HB star. The
hottest HB stars with envelope masses below 0.02M$_\odot$ do not have
active hydrogen-burning shells (T$_{\rm eff} >$23,000~K, extreme HB
stars = EHB stars\footnote{EHB stars show up as subdwarf B stars in
the field of the Milky Way and are considered good candidates for the
cause of the UV excess in elliptical galaxies.}). The increase in
bolometric correction with increasing temperature turns the horizontal
branch into a vertical blue tail (in optical colour-magnitude
diagrams) with the faintest blue tail stars being the hottest and
least massive ones.

In the optical colour-magnitude diagram the blue hook stars show up at
the very faint end of the blue tail in agreement with their high
temperatures suggested by the UV photometry. They populate a range in
visual magnitude (and thus effective temperature) beyond that of the
hottest EHB stars analysed in globular clusters so far (NGC 6752, Moehler
et al. 2000b), which already populate the canonical EHB to its very
hot end. Even hotter EHB stars cannot be produced by simply reducing
their envelope mass, since a minimum mass of the hydrogen envelope is
required for canonical models to initiate the helium core flash at the
tip of the red giants branch (see, e.g., Brown et al. 2001).
Obviously other evolutionary channels are needed to produce the blue
hook stars.

The concept of the delayed helium flash has been first suggested by
Castellani \& Castellani (1993). In this scenario a star loses so much
mass on the red giant branch (RGB) that it fails to ignite helium core
burning on the RGB and instead starts to evolve into a helium core
white dwarf. During this evolution it may, however, still ignite
helium burning in its core. Depending on when this happens one
distinguishes between ``early'' (flash at the hot top of the white dwarf
cooling curve) and ``late'' hot flashers (flash along the white dwarf
cooling curve). Cool flashers in this diction are stars igniting
helium core burning on the RGB. D'Cruz et al. (1996, 2000) discussed
in detail the evolution of early hot flashers in connection with the
blue hook stars. The evolution of such an early hot flasher, however,
proceeds rather similar to that of a canonical EHB star once it has
settled down to stable helium core burning. It can therefore not
explain the faint UV luminosities and high expected temperatures of
the observed blue hook stars.

Brown et al. (2001) studied also the case of a late hot flasher which
evolves quite differently: If the flash takes place late enough along
the white dwarf cooling curve the hydrogen burning shell has
weakened sufficiently to allow mixing between the helium core and the
hydrogen envelope during the helium core flash. Thus hydrogen is mixed
towards the interior of the star and mostly burned to helium, while
the surface is highly enriched in helium and some carbon/nitrogen. Due
to the lack of hydrogen in the star's atmosphere absorption in the
extreme UV is considerably reduced and thus less flux is emitted at
far UV wavelengths, producing stars that lie below the canonical
zero-age EHB in UV colour-magnitude diagrams. In addition the late hot
flashers should lie at higher effective temperatures
($\approx$37,000~K) and -- due to the small difference in mass loss
between early and late hot flashers -- a gap should separate them from
the early hot flashers and canonical EHB stars at and below
$\approx$31,000K. Such a gap is indeed observed in NGC~2808 (Walker
1999).

One of the easiest verifications of the late hot flasher scenario is
the determination of temperatures and surface compositions for
blue hook stars. The distinction between helium-rich late hot flashers
and canonical EHB stars is facilitated by the diffusion active in EHB
stars that makes their atmospheres helium-poor (e.g. Moehler et
al. 2000b, see also review by Bradford Behr in this volume) 
with helium abundances generally below 0.1 solar.

Moehler et al. (2002) observed medium resolution spectra of blue hook
stars with the New Technology Telescope of ESO.  Indeed the stars
turned out to be helium-rich with 9 of 12 stars showing at least solar
helium abundance (i.e. more than a factor of 10 higher than the
hottest canonical EHB stars observed in NGC~6752) and four even show
helium abundances of Y$>0.7$. There are also indications of
super-solar carbon abundances, which would be consistent with
predictions of the late hot flasher scenario. However, contrary to
expectation the stellar atmospheres still contain quite some
hydrogen. This may be understood by the results of Schlattl \& Weiss
(priv. comm.), who found that a small amount of hydrogen
survives the flash mixing. While diffusion can enrich the atmospheres
of the hot flashers in hydrogen and deplete them in helium more
detailed calculations will be necessary to verify if this process is
efficient enough to explain the observations. Recent observations with
the ESO VLT of the blue hook stars in NGC~2808 also show helium-rich
spectra, thus supporting the late hot flasher scenario, but no
detailed analysis has been done so far.

\section{The Real Thing}
As white dwarfs are the final stage in the evolution of all low mass
stars many are expected to exist in globular clusters. Their faintness,
however, allowed them to escape detection until 1995: That year not
only saw detections of a few stars that could be white dwarfs in M~15
(de Marchi \& Paresce 1995) and $\omega$ Cen (Elson et al. 1995),
but also of prominent white dwarf sequences in NGC~6397 (Paresce et
al. 1995) and M~4 (Richer et al. 1995). Later observations managed to
hunt down white dwarfs also in NGC~6752 (Renzini et al. 1996) and
47~Tuc (Zoccali et al. 2001).

\subsection{The bright end -- Nonflickerers}
Cool et al. (1998) detected several faint but UV bright stars in
NGC~6397. While some of those showed light variations and colours 
consistent with being
cataclysmic variables three remained blue in all filters and did not
show any variations. This earned them the name ``nonflickerer'' and
their faintness suggested an identification with low-mass helium core
white dwarfs of about 0.25~M$_\odot$. A spectrum obtained by Edmonds
et al. (1999) showed a hydrogen-rich atmosphere and yielded parameters
consistent with a 0.25~M$_\odot$ white dwarf. The radial velocity
obtained from the spectrum deviated from that of the cluster
indicating a binary nature of the star, which would be consistent with
its low mass. 

\subsection{Distance determinations and spectral types}
Renzini et al. (1996) suggested to use the white dwarf sequence for
distance determinations in the same way as the main sequence: 
comparing the white dwarf sequence of a globular cluster to an
appropriate cooling sequence of local white dwarf with well determined
trigonometric parallaxes yields the distance of the globular
cluster. While it may seem strange to use the faintest objects in a
globular cluster to derive its distance white dwarfs offer
considerable advantages as standard candles when compared to main
sequence stars:
\begin{itemize}
\vspace*{-1ex}
\item They come in just two varieties - either hydrogen-rich (DA) or
helium-rich (DB) {\em independent of their original metallicity}. So
the problem of finding local stars of the same metallicity as the
metal-poor globular clusters vanishes.
\vspace*{-1ex}
\item White dwarfs are locally much more numerous than metal-poor main
sequence stars allowing to define a better reference sample.
\vspace*{-1ex}
\end{itemize}
However, there are also some problems (for details see Salaris et
al. 2001). The brightness of white dwarfs depends on their
\begin{itemize}
\item {\bf mass}\\
There are so far no observational mass determinations available for any white
dwarfs in globular clusters.
Renzini et al. (1996) argue on theoretical grounds
for a mean mass of 0.53$\pm$0.02~M$_\odot$
for white dwarf in globular clusters. Unfortunately, there are {\em
no} local white dwarfs in this mass range with directly determined
masses. All masses for local white dwarfs in this range are derived
from effective temperatures and surface gravities in combination with
evolutionary models. Masses derived this way depend on the assumed mass of
the remaining hydrogen layer -- changing the mass of this layer from 0 to
10$^{-4}$~M$_\odot$ changes the resulting white dwarf mass by 0.04~M$_\odot$.
\item {\bf spectral type}\\ 
DB stars are fainter than DA stars at a given colour, with the offset
depending on the filter combination (i.e. the offset is greater in $V$
vs. $B-V$ than in $I$ vs. $V-I$). However, more massive stars are also
fainter at a given colour and the best way to verify the spectral
types of white dwarfs is -- obviously -- by spectroscopy. This has
been done with VLT observations of 
white dwarfs in NGC~6397 (Moehler et al. 2000a), M~4 and
NGC~6752 (see Fig.~2). All observed candidates turned out
to be hydrogen-rich DA white dwarfs as was assumed for the distance
determinations of NGC~6752 (Renzini et al. 1996) and 47~Tuc (Zoccali
et al. 2001).
\end{itemize}
\begin{figure}
\plotfiddle{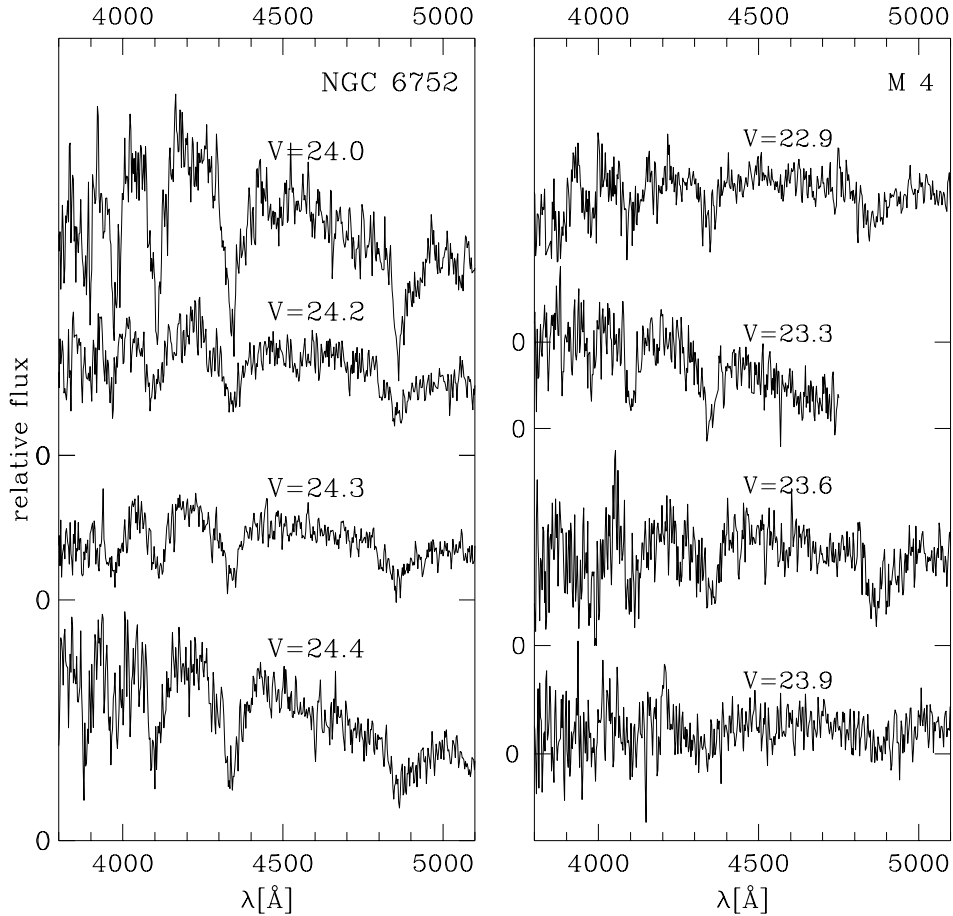}{11.7cm}{0}{64}{64}{-190}{-100}
\caption{VLT spectra of white dwarfs in M~4 (ESO proposal
65.H-0531(A), 1.5 hours exposure time) and NGC~6752 (ESO proposal
67.D-0201(B), 9 hours exposure time). All spectra show the strong
Balmer lines typical for DA white dwarfs. More details will be given
in Moehler, Heber, Napiwotzki, Koester \& Renzini (in prep.)}
\end{figure}

The distance moduli derived from white dwarfs for NGC~6752 and 47~Tuc
are in good agreement with other methods, esp. since the distance to
47~Tuc derived from main sequence fitting has been shortened recently
by Percival et al. (2002). 
\subsection{The faint end -- age determination}
Recent very deep HST observations allowed to detect the white dwarf
cooling sequence in M~4 to unprecedented depths of $V\approx$30
(Richer et al. 2002). Using proper motions to separate field and
globular clusters stars (a method first used by King et al. 1998 for
NGC~6397) the globular cluster sequences were isolated from
field stars. The luminosity function of the globular cluster white
dwarfs shows a much more sudden increase
towards fainter magnitudes than that of the disk white dwarfs, suggesting that
star formation in M~4 took place over a very short time (Hansen et
al. 2002). 
Assuming that the photometry really reaches the end of the white
dwarf sequence one can derive the age of the globular cluster from the age
of its oldest and faintest white dwarfs. Aside from the observational
uncertainty also uncertainties in the cooling tracks affect the result
(see Chabrier et al. 2000 for more details). As a preliminary result
Hansen et al. (2002) derive an age of 12.7$\pm$0.7 Gyr from the white
dwarf luminosity function of M~4, consistent with other independent age
estimates.

\acknowledgements I am grateful to the staff people at ESO for making
most of my work cited here possible. I also want to thank my
collaborators U. Heber, D. Koester, W.B. Landsman, R. Napiwotzki,
A. Renzini, and A.V. Sweigart for the good team work that enabled us
to get exciting results. 

\end{document}